\documentclass[fleqn,twoside]{article}%
\usepackage{amsmath}
\usepackage{amsfonts}
\usepackage{amssymb}
\usepackage{graphicx}
\usepackage{cite}
\def\be{\begin{equation}}
\def\ee{\end{equation}}
\def\bi{\bibitem}
\begin{document}
\title{Bianchi II, VIII, and IX viscous fluid cosmology.}
\author{A. Banerjee$^1$, Abhik Kumar Sanyal$^2$ and S. Chakraborty$^3$}
\maketitle
\noindent
\begin{center}
\noindent
$^{1,2}$ Dept of Physics, Jadavpur University, Calcutta 700 032, India.\\
$^3$ Dept of Mathematics, Jadavpur University, Calcutta 700 032, India.
\end{center}
\footnotetext{\noindent
Electronic address:\\
\noindent
$^1$ asit@juphys.ernet.in\\
$^2$ sanyal\_ ak@yahoo.com;
Present address: Dept. of Physics, Jangipur College, India - 742213.\\
$^3$ schakraborty@math.jdvu.ac.in}
\noindent
\abstract{In this paper we study the exact solutions for a viscous fluid distribution in Bianchi II, VIII, and IX models. The metric is simplified by assuming a relationship between the coefficients and the metric tensor. Solutions are obtained in two special cases: in one an additional assumption is made where the matter density and the expansion scalar have a definite relation and in the other a barotropic equation of state of the form $p = \epsilon\rho$ is assumed. While the Bianchi II solutions are already found in the literature the other two classes of solutions are apparently new.}
\maketitle
\flushbottom
\section{Introduction:}

Cosmological principle states that the universe was initially isotropic and homogeneous. This principle may be sacrificed to consider that there was a large amount of anisotropy at the very early universe. However, the present universe is isotropic at large scale. One of the possible mechanisms to decrease the initial anisotropy is to consider the existence of dissipative phenomena such as the viscosity \cite{1}. This also results in a large increase of entropy per baryon as observed in the present-day universe \cite{2, 3, 4}. Considering the possible importance of the dissipative phenomena in the evolution of the cosmological model it is worth to attempt exact solutions which should at the same time be physically reasonable. This was attempted earlier by several authors considering different Bianchi models \cite{5, 6, 7, 8, 9, 10}. Bianchi types - II, VIII, and IX space-times with a viscous fluid were previously considered in \cite{5}. They obtained linear differential equations in terms of one of the metric coefficients imposing certain restrictions on the viscosity parameters as well as the geometric scalars like shear and expansion. Exact solutions could not, however be explicitly given. Considering the possible importance of the dissipative phenomena in the evolution of the cosmological models it is worthwhile to attempt towards obtaining exact solution which may at the same time be physically reasonable. In the present paper such solutions are obtained for the three Bianchi  varieties mentioned above starting from the special form of the metric used previously \cite{5, 11}. Considering the number of field equations available and the list of unknowns, two different restrictions were imposed. In addition to a relation assumed between the two metric coefficients there was an assumption connecting matter density and the expansion scalar in one class of solutions and a barotropic equation in the other class. The Bianchi - II solutions are found to be identical with those previously obtained \cite{7}, whereas those obtained for Bianchi VIII and IX appear to be new.\\

In the following section, Einstein's Field Equations for Bianchi - II, VIII, and IX have been presented and and Some some general results are explored. In section 3, exact solutions are found and from the long list of possible solutions, only those which are physically reasonable to some extent are given explicitly.

\section{Einstein's Field Equations and Some General Results:}

The orthogonal, locally rotationally-symmetric (LRS) metric for the spatially-homogeneous Bianchi-II ($\delta = 0$), VIII ($\delta = - 1$), and IX ($\delta = 1$) models can be written as \cite{5, 7}

\be\label{2.1}\begin{split} ds^2 = & -dt^2 + S^2 dx^2 + R^2\big[dy^2 + f^2(y) dz^2\big] - S^2 h(y) \big[2 dx - h(y) dz\big]dz,~~\mathrm{where,}\\&
f(y) = \sin y, ~~~~~~h(y) = \cos y,~~~~~~~ \mathrm{for}~~\delta = 1,\\&
f(y) = y, ~~~~~~~~~~h(y) = -{1\over 2}y^2,~~~~~~ \mathrm{for}~~\delta = 0,\\&
f(y) = \sinh y, ~~~~h(y) = -\cosh y,~~~ \mathrm{for}~~\delta = -1,\end{split}\ee
and, $R = R(t)$, $S = S(t)$, and the proper volume is $\sqrt{-g} = SR^2 f(y)$. The energy-momentum tensor for a viscous fluid is given by

\be\label{2.2}\begin{split}& T_{ij} = (\rho + \bar p )v_i v_j + \bar p g_{ij} - \eta U_{ij},\\& \mathrm{where},~ \bar p = p + {2\over 3} \eta {v^a}_{;a},~\mathrm{and},~~~ U_{ij} = v_{i;j} + v_{j;i} + v_i v^a v_{j;a} + v_j v^a v_{i;a}, ~~\mathrm{with},~~~v_i v^j = -1.\end{split}\ee
In the above, $\rho$, $p$, and $\eta$ are the mass density, thermodynamic pressure, and coefficient of shear viscosity, respectively, while $v_i$ is the four-velocity. In the system of units $8\pi G = c = 1$, Einstein's field equations

\be\label{2.3} {R_i}^j = {1\over 2}{\delta_i}^j R = - {T_i}^j,\ee
in the co-moving coordinate system $v^i = {\delta_0}^i$, can be written, in view of Equations \eqref{2.1} and \eqref{2.2} as

\be\label{2.4}2{\dot R\over R}{\dot S\over S} + {\dot R^2 + \delta \over R^2} -{1\over 4}{S^2\over R^4} = \rho,\ee
\be\label{2.5} {\ddot R\over R}+{\ddot S\over S}+{\dot R\over R}{\dot S\over S}+{1\over 4}{S^2\over R^4} = - p +{2\over 3}\eta\left({\dot R\over R} - {\dot S\over S}\right),\ee
\be\label{2.6} 2{\ddot R\over R} + {\dot R^2 + \delta \over R^2} + {3\over 4}{S^2\over R^4} = -p - {4\over 3}\eta\left({\dot R\over R} - {\dot S\over S}\right),\ee
where a dot indicates differentiation with respect to time. The expansion ($\theta$) and shear ($\sigma^2$) scalars have usual definitions: i.e.,

\be\label{2.61}\begin{split} &\theta = {v^i}_{;i}, ~~~\mathrm{and},~~~\sigma^2 =\sigma_{ij}\sigma^{ij},\\& \sigma_{ij} = v_{(i;j)} +{1\over 2}\big(v_{i;a}v^a v_j + v_{j;a}v^a v_i\big) - {1\over 3}\theta \big(g_{ij} + v_i v_j\big).\end{split}\ee
In view of these relations we find that

\be\label{2.7}\theta = 2{\dot R\over R} + {\dot S\over S}~~~\mathrm{and},~~~\sigma^2 = {2\over 3}\left({\dot R\over R} - {\dot S\over S}\right)^2.\ee
Furthermore, from Equations \eqref{2.7} one can also find that

\be\label{2.8} 3{\dot R\over R} = \theta \pm \sqrt{3\over 2}~ \sigma,~~~~~3{\dot S\over S} = \theta \mp \sqrt{6}~\sigma.\ee
As a consequence of Bianchi identity ${{T_i}^j}_{;j} = 0$ we get

\be\label{2.9} \dot\rho = -(\rho + p) \theta + 4\eta \sigma^2.\ee
The Raychaudhuri equation can be written as

\be\label{2.10}\begin{split}& \dot\theta = -{1\over 3}\theta^2 - 2\sigma^2 + R_{ij} v^i v^j,\\&\mathrm{where,}~~R_{ij} v^i v^j = -{1\over 2}\big(\rho + 3p\big).\end{split}\ee

\section{Exact Solutions of Einstein's Equation:}

Our aim is to find the exact solutions of the field equations \eqref{2.4} - \eqref{2.6} having five unknown functions: viz., $\rho,~ p,~\eta,~ R,~ \mathrm{and}~ S$. In order to remove two extra degrees of freedom we require a pair of physically reasonable equations. Our first assumption relates the metric coefficients $R$ and $S$ in the form

\be\label{3.1} S = \mu R^k,\ee
where $\mu$ and $k$ are arbitrary constants. Equation \eqref{3.1} yields ${\dot S\over S} = k{\dot R\over R}$, which in turn used in \eqref{2.7} to yield

\be\label{3.2} \theta = (k + 2){\dot R\over R},~~~\mathrm{and}, ~~~\sigma^2 = {2\over 3}\big(k - 1\big)^2{\dot R^2\over R^2}.\ee
We note, however, that in view of \eqref{3.2}, $\theta$ and $\sigma^2$ are related by

\be\label{3.3}\begin{split} &\sigma^2 = D^2 \theta^2,\\& \mathrm{where},~~D^2 = {3\over 2}\left[{k+2\over k-1}\right]^2,\end{split}\ee
which is a constant. Now by virtue of Equation \eqref{3.1}, the field equations \eqref{2.4} to \eqref{2.6} take the forms:

\be\label{3.4} (2k + 1){\dot R^2\over R^2} + {\delta\over R^2} - {\mu^2\over 4} R^{2k-4} = \rho,\ee

\be\label{3.5} (k+1) {\ddot R\over R} + k^2 {\dot R^2\over R^2} + {\mu^2\over 4}R^{2k-4} = -p + {2\over 3}(1-k){\dot R\over R}\eta,\ee

\be\label{3.6} 2{\ddot R\over R} + {\dot R^2 + \delta\over R^2} + {3\over 4}\mu^2 R^{2k-4} = -p - {4\over 3}(1-k){\dot R\over R}\eta.\ee
From Equations \eqref{3.4} and \eqref{3.6} we can obtain two other useful relations - viz.,

\be\label{3.7} (\rho + 3p) = -2(k+2){\ddot R\over R} - 2k(k-1){\dot R^2\over R^2},\ee
and

\be\label{3.8} 2\eta(1-k){\dot R\over R} = (k-1){\ddot R\over R} + (k^2 -1) {\dot R^2\over R^2} + \mu^2 R^{2k-4} -{\delta\over R^2}.\ee
Hence, once $R$ is known as a function of time, one can use Equation \eqref{3.2} to find $\theta$ and $\sigma^2$. $\rho$ can be obtained from \eqref{3.4}, $p$ from \eqref{3.7}, and finally $\eta$ can be found from Equation \eqref{3.8}.\\

\subsection{Case 1:}

Here we attempt solutions under the assumption

\be\label{3.9} \rho = C^2 \theta^2,\ee
where $C$ is a constant. Now in view of Equations \eqref{3.9}, \eqref{3.2}, and \eqref{3.4} we get

\be\label{3.10} \dot R^2 = {\mu^2 \over 4k_1} R^{2(k-1)} - {\delta\over k_1},\ee
where the constant $k_1 = 2k + 1 - C^2(k + 2)^2$. Now considering $k_1$ positive, the differential equation \eqref{3.10} can be written in the following integral form

\be\label{3.11} \int{dR\over \left[R^{2(k-1)} - {4\delta\over \mu^2}\right]^{1\over 2}} = k_2(t - t_0),\ee
where ${k_2}^2 = {\mu^2\over 4k_1}$ is a constant and $t_0$ is the constant of integration. The above integral can be evaluated for $k = 0, {1\over 2}, {3\over 2}, 1, ~\mathrm{and}~ 2$. However, we shall consider only those cases for which the solutions are physically reasonable - viz., $\rho > 0$, $p > 0$, $\eta > 0$, etc. These are the cases for $k = 0,~{1\over 2}~ \mathrm{and} ~{3\over 2}$.\\

\noindent
\textbf{a. Solution for $k = 0$:}\\

\noindent
The solutions for the Bianchi-IX model $(\delta = + 1)$ are

\be\label{3.12}\begin{split}& R^2 = {\mu^2\over 4} - {(t_0-t)^2\over k_1} ;\hspace{0.1 in} S = \mu; \hspace{0.1 in}\theta = {2\over k_1}\left[{t_0-t\over {\mu^2\over 4}-{{(t_0-t)^2}\over k_1}}\right]; \hspace{0.1 in}\sqrt{-g} = \left[{\mu^2\over 4} - {(t_0-t)^2\over k_1}\right]\sin{y};\\&
\sigma^2 = D^2 \theta^2; \hspace{0.2 in}\rho = C^2 \theta^2;\hspace{0.2 in}p = {4\over 3k_1}\left[{{\mu^2\over 4} - {C^2\over k_1}(t_0-t)^2\over \left({\mu^2\over 4} - {(t_0-t)^2\over k_1}\right)^2}\right];\hspace{0.1 in}\eta = {1\over 2(t_0 -t)}\left[k_1\left({{3\mu^2\over 4} + {(t_0-t)^2\over k_1}\over {\mu^2\over 4} - {(t_0-t)^2\over k_1}}\right)^2+1\right].\end{split}\ee
Thus $S$ does not evolve and the above solution implies that

\be\label{3.121} {\mu^2\over 4} - {(t_0-t)^2\over k_1} > 0.\ee
Furthermore, since  $k_1 > 0$, the model is expanding for $t < t_0$. At the initial epoch $t = t_0 - {1\over 2}\mu\sqrt{k_1}$, the universe starts from a zero proper volume, while $\theta,~\sigma^2,~\rho,~p,~\mathrm{and}~\eta$ have infinitely large magnitude, and the universe expands with increasing $t$ till $t = t_0$. The proper volume tends to ${\mu^3\over 4}\sin {y}$, a finite value, while $\theta,~\sigma^2,~\rho,~\eta$ all vanish. Subsequently there is a reversal of the motion, i.e., $\theta < 0$ allowing a contracting model. However, during contraction $\eta < 0$ throughout, which is not a physical situation.\\

The solutions for the Bianchi II ($\delta = 0$) model are

\be\label{3.13}\begin{split} &R^2 = 2k_2(t-t_0);\hspace{0.25 in}S = \mu; \hspace{0.25 in}\sqrt{-g} = 2\mu k_2(t-t_0)y;\hspace{0.25 in} \theta = (t-t_0)^{-1};\\&\sigma^2 = D^2 \theta^2;\hspace{0.25 in}\rho = C^2 \theta^2; \hspace{0.25 in}p = {1-C^2\over 3} k_2(t-t_0)^{-2};\hspace{0.25 in} \eta = {\mu^2\over 4k_2^2}k_2(t-t_0)^{-1}.\end{split}\ee
In this case $\rho$ and $p$ are linearly related indicating a barotropic equation of state. These solutions are quite in agreement with those obtained previously by Banerjee et at., \cite{7}. We shall not present the solutions for the Bianchi-VIII ($\delta = - 1$) model, since in this case pressure becomes negative at some stage.\\

\noindent
\textbf{b. Solution for $k = {1\over 2}$:}\\

\noindent
For $k = {1\over 2}$, solutions for Bianchi-VIII and IX models are not obtained in a closed form. So we shall present the solutions for the Bianchi-II model only. With $k = {1\over 2}$ Equation \eqref{3.11} can be integrated to yield

\be\label{3.14}\begin{split} &R^2 = \left[{3\over 2}k_2(t-t_0)\right]^{4\over 3};\hspace{0.1 in}S^2 = \mu^2\left[{3\over 2}k_2(t-t_0)\right]^{2\over 3}; \hspace{0.1 in} \sqrt{-g} = \mu\left[{3\over 2}k_2(t-t_0)\right]^{5\over 3}; \hspace{0.1 in} \theta ={5\over 3}(t-t_0)^{-1};\\&
\sigma^2 = D^2 \theta^2;\hspace{0.25 in} \rho = C^2 \theta^2; \hspace{0.25 in} p = \left[{12-25 C^2\over 9}\right](t-t_0)^{-2};\hspace{0.25 in} \eta = \left[{8k - 1\over 3}\right](t-t_0)^{-1}.\end{split}\ee

\subsection{Case 2:}

Solutions satisfying the equation of state

\be\label{3.15} p = \epsilon\rho;\hspace{0.2 in} \mathrm{with}\hspace{0.2 in} 0\le \epsilon \le 1.\ee
Equations \eqref{3.4} and \eqref{3.7} together with the above assumption lead to the differential equation

\be\label{3.16}\begin{split}& 2R^a \ddot R + a R^{(a-1)}\dot R^2+ b\delta R^{(a-1)} - C R^{(2k-a-3)}=0,\\&
\mathrm{where~ the~ constants}~ a,~ b,~ c ~\mathrm{are~ given~ as~}\\&
a = {2k^2 + 1 + 3\epsilon(2k + 1)\over k+ 2},\hspace{0.2 in}b = {1 + 3\epsilon\over k + 2},\hspace{0.2 in} c = {\mu^2(1+ 3\epsilon)\over 4(k+2)}.
\end{split}\ee
We note that $a,~ b,~ c$ are all positive for $k > 0$. If we integrate Equations (3.16) twice, we find that

\be\label{3.17} {dR\over \left[ l_0 R^{-a} + {C\over 2k+a-2}R^{2(k-1)}-{b\delta\over a}\right]^{1\over 2}} = t - t_0,\ee
where $l_0$ and $t_0$ are constants of integration. We set $l_0 = 0$ and observe that the integral can be evaluated for $k = 0,~ {1\over 2},~ {3\over 2},~ \mathrm{and}~ 2$. However, physically reasonable solutions are available for $k = {3\over 2}~ \mathrm{and} ~2$, which are described below.\\

\noindent
\textbf{a. Solution for $k = {3\over 2}$:}\\

\noindent
The integral \eqref{3.17} now can at once be evaluated, and the solutions can be obtained as before. For Bianchi - IX ($\delta = 1$), the solutions are

\be\label{3.18}\begin{split} & R = n^2 + {l^2\over 4}(t-t_0)^2;\hspace{0.2 in}S = \left[n^2 + {l^2\over 4}(t-t_0)^2 \right]^{3\over 2};\hspace{0.2 in}\sqrt{-g} =\mu \left[n^2 + {l^2\over 4}(t-t_0)^2 \right]^{7\over 2} \sin{y}\\&
\theta = {7l^2\over 4}\left[{t-t_0\over n^2 + {l^2\over 4}(t-t_0)^2}\right];\hspace{0.2 in}\eta = {2\over l^2(t-t_0)}\left[{1-n^2\big({l^2\over 4}+\mu^2\big)-{l^2\over 4}\big({3l^2\over 2}+\mu^2\big)(t-t_0)^2\over n^2 + {l^2\over 4}(t-t_0)^2}\right]\\&
\sigma^2 = D^2\theta^2;\hspace{0.2 in}\rho = {l^2\big(l^2 - {\mu^2\over 16}\big)(t-t_0)^2 + 1 -{\mu^2 n^2\over 4}\over\big[ n^2 + {l^2\over 4}(t-t_0)^2\big]^2};\hspace{0.2 in}p = \epsilon \rho,
\end{split}\ee
where $l^2,~m^2~\mathrm{and}~n^2$ are constants. The above solutions \eqref{3.18} allow both contracting and expanding models for the universe. At a very early epoch $t\rightarrow \infty,~\sqrt{-g},~R,~S$ all were infinitely large, while, as one expects, $\theta,~\sigma^2,~\rho,~p,~\eta$ were insignificant. The universe contracts and at $t\rightarrow t_0,~\sqrt{-g},~R,~S$ take finite magnitude together with $\rho$ and $p$ while $\theta$ and $\sigma^2$ vanish but $\eta\rightarrow\infty$. For $t > t_0$ one gets expansion and at the final stage of solution, i.e., as $t\rightarrow \infty, ~\sqrt{-g},~R$ and $S$ again become infinitely large while the scalars $\theta,~\sigma^2,~\rho$ and $p$ vanish. However, at this stage the shear viscosity coefficient becomes negative.\\

The solutions for the Bianchi - II ($\delta = 0$) model are

\be\label{3.19}\begin{split} & R= {l^2\over 4}(t-t_0)^2;\hspace{0.2in}S = \mu\left[{1\over 3}l (t-t_0)\right]^3;\hspace{0.2in}  \sqrt{-g} =\left[{1\over 2}l(t-t_0)^7\right]y;\hspace{0.2in}\theta =7(t-t_0)^{-1};\\& \eta = -\left({2\mu^2\over l^2} + 3\right)  (t-t_0)^{-1};\hspace{0.2in}\sigma^2 = D^2\theta^2;\hspace{0.2in} \rho = \left(16-{\mu^2\over l^2}\right)(t-t_0)^{-2};\hspace{0.2in} p = \epsilon \rho.\end{split}\ee
In the above solutions, we find $\eta> 0$ demands $t < t_0$, which implies $\theta < 0$. So the solutions admit of a collapsing universe. However, to keep $\sqrt{-g} > 0$, we consider $l < 0$ for this case. Our universe starts collapsing from $t\rightarrow \infty$ when $R,~ S$ and $\sqrt{-g}$ take indefinitely large magnitude while $\theta,~\sigma^2,~\rho,~p,~\eta$ are insignificantly small. As $t$ increases, i.e., $t\rightarrow t_0$, proper volume vanishes along with metric coefficients $R$ and $S$, but this time $\theta \rightarrow \infty$ while $\sigma^2,~\rho,~p,~\eta$, all tend to infinity. This is a singular state.\\

Finally, the solutions for the Bianchi VIII ($\delta = -1$) model a

\be\label{3.20}\begin{split} &
R= {l^2\over 4}(t-t_0)^2 - n^2;\hspace{0.2in}S = \mu\left[{l^2\over 4} (t-t_0)^2 - n^2\right]^{3\over 2};\hspace{0.2in}  \sqrt{-g} =\mu \left[{l^2\over 4}(t-t_0)^2 - n^2 \right]^{7\over 2}\sinh{y};\\&\theta ={7~l^2\over 4}\left[{(t-t_0)\over {l^2\over 4}(t-t_0)^2 - n^2}\right];\hspace{0.2in}
\eta = {2\over l^2(t-t_0)}\left[{n^2\big({l^2\over 4}+\mu^2\big) -1 -{l^2\over 4}\big({3\over 2}l^2 + \mu^2\big)(t-t_0)^2 \over \big[{l^2\over 4}(t-t_0)^2 - n^2\big]^2}\right]   \\&
\sigma^2 = D^2\theta^2;\hspace{0.2in}\rho = {l^2\big(l^2 -{\mu^2\over 16}\big)(t-t_0)^2 + {\mu^2 n^2\over 4}-1\over \big[{l^2\over 4}(t-t_0)^2 - n^2\big]^2};\hspace{0.2in} p = \epsilon \rho.
\end{split}\ee
The above solutions demand $\big[{l^2\over 4}(t-t_0)^2 -n^2\big] > 0$. So, for $t > t_0$ we get the
expanding model $\theta > 0$ and for $t < t_0$, the contracting model is observed where $\theta < 0$.
For an expanding model there is an initial singularity at time given by
${l^2\over 4}(t-t_0)^2 - n^2 =0$. At this epoch $\theta,~\sigma^2,~\rho$ are all infinitely large. One can ensure the positivity of the matter density $\rho$ by suitable choices of the constant parameters. As $t\rightarrow\infty$ one has all the scalars mentioned above vanishingly small. On the other hand, for $t < t_0$ one can replace ($t - t_0$) by $ -(t_0 - t)$ in the expressions \eqref{3.20} and finds that for $t \rightarrow -\infty$ the proper volume attains an infinitely large magnitude. The contraction continues till the model collapses to a singularity of zero volume at $(t - t_0)^2 = {4n^2\over l^2}$. However, as $t$ increases $\eta$ becomes negative, though at sufficiently large $t$ (as $t \rightarrow \infty$) $\eta \rightarrow 0$.\\

\noindent
\textbf{b. Solution for $k = 2$:}\\

\noindent
Again we integrate Equation \eqref{3.17} to find $R$ in different models, hence, the solutions are obtained as before. For the Bianchi - IX model ($\delta = 1$), the solutions are

\be\label{3.21}\begin{split} & R = n \cosh{[l(t-t_0)]};\hspace{0.2in}S = \mu n^2 \cosh^2{[l(t-t_0)]};\hspace{0.2in}\sqrt{-g} = \mu n^4 \cosh^4{[l(t-t_0)]}\sin{y};\\&\theta = 4l\tanh{[l(t-t_0)]};\hspace{0.2in}
\sigma^2 = D^2\theta^2;\hspace{0.2in}\rho = 5l^2\tanh^2{\Big[l(t-t_0)\Big]}+{1\over n^2}\sec\mathrm{h}^2{[l(t-t_0)]} -{\mu^2\over 4};\\&p = \epsilon\rho;\hspace{0.2in}
\eta = {1\over 2l\tanh{[l(t-t_0)]}}\left[{1\over n^2\cosh^2{[l(t-t_0)]}} -3 l^2\tanh^2{[l(t-t_0)]} - l^2 -\mu^2\right].\end{split}\ee
The above solutions with $l > 0$ have apparently a lower bounce at time given by $t = t_0$
where the expansion scalar vanishes. The proper volume is infinite for $t =\pm\infty$. So an initially contracting phase will reverse into an expansion at an instant $t = t_0$. The anisotropy $\sigma^2$ momentarily disappears at the turning point. The possibility of $\rho$, however, depends on the choice of the constant parameters, and $\eta$ is not positive throughout the evolution.\\

For the Bianchi - II model ($\delta = 0$), solutions are

\be\label{3.22}\begin{split} & R = \exp{[l(t-t_0)]};\hspace{0.2 in}S = \exp{[2l(t-t_0)]};\hspace{0.2 in} \sqrt{-g} = \exp{[4l(t-t_0)]y};\\& \theta = 4l;\hspace{0.2 in}
\sigma^2 = D^2 \theta^2;\hspace{0.2 in} \rho = 5l^2 -{\mu^2\over 4};\hspace{0.2 in} p = \epsilon\rho;\hspace{0.2 in}\eta = -{l\over 2}\big(4l^2+\mu^2\big).\end{split}\ee
In order to keep $\eta > 0$ throughout the evolution, we must have $l < 0$, which demands the model is contracting. The interesting feature of these solutions is that as the universe collapses with $\theta =$ constant, $R,~S,~\sqrt{-g}$ fall exponentially, while $\rho,~ p,~\sigma^2$ remain constant.\\

Finally, for the Bianchi - VIII model ($\delta = -1$), the solutions are

\be\label{3.23}\begin{split} & R = n \sinh{[l(t-t_0)]};\hspace{0.2 in}S = \mu n^2 \sinh^2{[l(t-t_0)]};\hspace{0.2 in} \sqrt{-g} = \mu n^4 \sinh^4{[l(t-t_0)]}\\&
\theta = 4l\coth{[l(t-t_0)]};\hspace{0.2in}
\sigma^2 = D^2\theta^2;\hspace{0.2 in}\rho = 5l^2\coth^2{[l(t-t_0)]} - {\mathrm{cosec}\mathrm{h}^2\over n^2}{[l(t-t_0)]}-{\mu^2\over 4};\\&p = \epsilon \rho;\hspace{0.2 in}
\eta = {\tanh{[l(t-t_0)]}\over 2l}\left[l^2 + \mu^2 +3l^2\coth^2{[l(t-t_0)]}\right] + {\mathrm{cosec}\mathrm{h}^2{[l(t-t_0)]}\over n^2}.\end{split}\ee
Here again, positivity of $\eta$ is ensured when $l < 0$ and $\sinh{[l (t - t_0)]} > 0$. In this case $\theta < 0$ that is there is only a contracting model. At $t\rightarrow \infty$, the proper volume is infinitely large since the contraction rate being also indefinitely large. As $t$ approaches $t_0$ we have a point singularity both $R$ and $S$ vanishing but $\theta$ remaining negative and finite.

\end{document}